\documentclass[aps,pre,onecolumn,superscriptaddress,groupedaddress]{revtex4}

\usepackage[utf8]{inputenc}
\usepackage[english]{babel}
\usepackage{indentfirst}
\usepackage{fancyhdr}
\usepackage[final]{showkeys}
\usepackage{graphicx}
\usepackage{subfig}
\usepackage{amssymb}
\usepackage{comment}
\usepackage{latexsym}
\usepackage{amsthm}
\usepackage{amsmath}
\usepackage{amsfonts}
\usepackage{mathtools}
\usepackage{esvect}
\usepackage{pgf,tikz}
\usepackage{mathrsfs}
\usepackage{epstopdf}
\usetikzlibrary{arrows}
\usepackage{enumerate}
\usepackage{epigraph}
\usepackage[a4paper, left=1.5cm, right=1.5cm, top=2cm, bottom=2cm]{geometry}
\usepackage{floatpag}
\usepackage{comment}

\citeindextrue

\newcommand{\braket}[1]{\ensuremath{\left\langle #1 \right\rangle}}

\begin{document}

\title{
High-precision 
anomalous dimension of $3d$ percolation 
from 
giant cluster slicing 
}

\author{Alessandro Galvani}
\affiliation{SISSA and INFN, Sezione di Trieste, Via Bonomea 265, I-34136 Trieste, Italy}

\author{Andrea Trombettoni}
\affiliation{Department of Physics, University of Trieste, Strada Costiera 11, I-34151 Trieste, Italy}
\affiliation{SISSA  and  INFN,  Sezione  di  Trieste, Via Bonomea 265, I-34136 Trieste, Italy}
\affiliation{CNR-IOM DEMOCRITOS Simulation Center and SISSA, Via Bonomea 265, I-34136 Trieste, Italy}

\author{Giacomo Gori}
\affiliation{Institut f\"ur Theoretische  Physik,  Universit\"at Heidelberg, D-69120  Heidelberg,  Germany}
\affiliation{CNR-IOM DEMOCRITOS Simulation Center and SISSA, Via Bonomea 265, I-34136 Trieste, Italy}

\begin{abstract} We apply the critical geometry approach for bounded critical phenomena 
\cite{gori} to $3d$ 
percolation. 
The functional shape of the order parameter profile $\phi$ 
is related via the fractional Yamabe equation to its scaling dimension $\Delta_{\phi}$.
We obtain $\Delta_{\phi}= 0.4785(7)$ from which the anomalous dimension $\eta$ is found to be $\eta=-0.0431(14)$, a value compatible with, and more precise than, its previous direct 
measurements. 
A test of 
hyperscaling is also performed.


\end{abstract}

\maketitle


\section{Introduction}Percolation takes a special spot 
among physical phenomena, describing the addition 
of sites or links to a bounded system and the 
formation at a critical threshold of a macroscopic cluster 
connecting the boundaries. Despite its 
simplicity, due to its deep geometrical meaning, percolation theory can model vastly different real-world phenomena \cite{Stauffer2018}, ranging from water passing through coffee to small molecules branching to form a gel~\cite{Flory1941,Stockmayer1944} or from wildfires~\cite{Caldarelli2001} to the spreading of infections~\cite{Grassberger,Sander2002,Miller2009}. 

Percolation provides a clear introduction to critical phenomena, with an easily identifiable transition and a visually striking example of self-similarity at the critical point \cite{Binney1992}. It differs from spin systems as it lacks a Hamiltonian, making it an easy to simulate, purely geometrical model. 
It has been the subject of several physical studies, via methods such as the renormalization group \cite{Reynolds1977,Hu1992,Rammal}, as well as boasting
a long history of 
mathematical investigations \cite{Grimmett1999}, culminating in a Fields medal \cite{Smirnov2001,Smirnov2007}. The richness of the field is reflected in the variety of related critical behaviors and universality classes, including 
directed \cite{Hinrichsen2000,Odor2004} and long-range \cite{Grassberger83,Aizenman1986,Newman02,Gori17} percolation, 
and by the different models within the same class, such as bond or site percolation. Remarkably, several exact results, including critical exponents, are available in two dimensions \cite{exp2d1,exp2d2}.

At the critical point, percolation is described by a logarithmic 
conformal field theory \cite{Cardy1999}, with a single primary field \cite{amit1977}. 
Bond percolation can be obtained as the analytic continuation of the $q$-state Potts model for $q\rightarrow 1$ \cite{pottsperc}. This procedure, however, does not preserve unitarity. For this reason,
the most accurate technique currently available to obtain critical exponents for $O(N)$ models, the conformal bootstrap \cite{bootstrap0}, cannot be straightforwardly applied to percolation, meaning that current results for anomalous dimensions are not especially precise. 
An updated list of values for percolation threshold and critical exponents is in \cite{wiki1,wiki2}.

In this work, we 
apply the geometric theory of bounded critical phenomena introduced in \cite{gori} to the case of $3d$ percolation. 
To extract more accurate critical profiles, 
we use the continuum percolation model. The procedure is based on slicing the percolating (giant) cluster emerging at the critical point, to then measure the fraction of 
the giant cluster at a given distance from the boundaries. 
We start by providing a brief summary of the critical geometry approach, then we discuss continuum percolation. After presenting the check done for $d=2$, we give our result for the scaling dimension $\Delta_{\phi}$ of the order parameter and the corresponding anomalous dimension $\eta$ of the $3d$ percolation. We conclude by checking 
hyperscaling relations with the obtained value of $\eta$.

\section{Critical geometry}The main property a system typically gains at its critical points is conformal invariance~\cite{Polyakov1970,Polyakov1974}. Heuristically, this means that every point and every region of the system look the same. 
Introducing a boundary clearly breaks this property. The question addressed in~\cite{gori} is then: is there a way to recover 
some degree of uniformity? 
If a metric is introduced which sets the boundary at an infinite distance, then there no longer is a distinction between points close to the boundary and points deep in the bulk. The 
changes of the euclidean metric that one can allow are 
pointwise scale changes,
 since the system must still be locally euclidean. This means that the choice of metric reduces to the choice of a function $\gamma(\mathbf{x})$, with $\mathbf{x}$ belonging to the considered bounded domain. $\gamma$ sets a local scale:
\begin{equation}\label{metrica}
\delta_{ij}\rightarrow g_{ij}=\frac{\delta_{ij}}{\gamma(\mathbf{x})^2},
\end{equation}
$\delta_{ij}$ being the flat metric, i.e.  the identity matrix, with $i,j=1,\ldots,d$.

Constraints on the function $\gamma$ have to be imposed. Since we have a curved space, we should look into the 
intrinsic quantities that describe its 
geometry, the most obvious one being the Ricci scalar curvature. The main hypothesis in \cite{gori} is that the metric must make a bounded critical system as uniform as possible: This means making the scalar curvature constant. This curvature would have to be negative, since spaces with positive curvature, like spheres, lack boundaries. The simplest examples of space with constant negative curvature are the Poincaré half plane and disk models. 


Starting from the metric $\eqref{metrica}$ with an unknown $\gamma(\mathbf{x})$, one can compute the Christoffel symbols $\Gamma_{jk}^i=\frac12 g^{il} \left( \partial_{k} g_{lj} + \partial_{j} g_{lk} -
 \partial_{l} g_{jk} \right)$, from which one gets the Ricci tensor
$\mathrm{Ric}_{ij} =
\partial_{l}{\Gamma^l_{ji}} - \partial_{j}\Gamma^l_{li}
+ \Gamma^l_{l\lambda} \Gamma^\lambda_{ji}
- \Gamma^l_{j\lambda}\Gamma^\lambda_{li}$,
and finally the Ricci scalar (summation over repeated indices is implied):
\begin{equation}
R=
\mathrm{Ric}_{ij} \, g^{ij}=\kappa, \qquad \kappa<0.
\end{equation}
 
Without losing generality, we can set $\kappa=-1$. 

Let us first consider models at their upper critical dimension $d=d_c$ \cite{gori,Galvani2021}, for which the anomalous dimension $\eta$ vanishes. We can write the requirement of constant scalar curvature as an equation for the factor $\gamma(\mathbf{x})$, obtaining what 
in geometry literature is called the Yamabe equation~\cite{Yamabe1960} ($\bigtriangleup$ is the laplacian in flat space):
\begin{equation}
(-\bigtriangleup)\gamma(\mathbf{x})^{-\frac{d-2}2}=-\frac{d(d-2)} 4 \gamma (\mathbf{x})^{-\frac{d+2}{2}}.
\label{YAM}
\end{equation}

With the condition $\gamma(\mathbf{x})=0$ at the boundaries
of the domain $\Omega$, one obtains solutions which, close to the boundary, are proportional to the euclidean distance from it. The distance from any point to the boundary, computed with this metric, is therefore infinite, as desired. 

The reason to introduce the Yamabe equation is that its solution allows us to determine correlation function once the function $\gamma(\mathbf{x})$
is used as a local gauge to measure distances. 
An operator $\phi$ is called a scaling operator
of dimension $\Delta_\phi$ if
its one-point correlation functions transform as $\braket{\phi_{\lambda\Omega}(\lambda \mathbf{x})}=\lambda^{-\Delta_{\phi}}\braket{\phi_{\Omega}(\mathbf{x})}$ under a dilation of the system $\Omega \rightarrow \lambda \Omega$. The scale factor transforms similarly: $\gamma_{\lambda\Omega}(\lambda \mathbf{x})=\lambda \gamma_{\Omega}(\mathbf{x})$.
Since $\gamma(\mathbf{x})$ is the only local length scale, 
one-point functions are determined in the critical geometry framework up to a constant $\alpha$:
\begin{equation}\label{1p}
\braket{\phi(\mathbf{x})}=
\frac{\alpha}{\gamma(\mathbf{x})^{\Delta_{\phi}}},
\end{equation}
where $\Delta_{\phi}=\frac{d-2}2$. By similar reasoning, one can obtain 
a prediction for 
two-point correlation functions \cite{gori}.



What we discussed so far is only valid for fields whose anomalous dimension $\eta$ vanishes, i.e. $\Delta_{\phi}=\frac{d-2}2$. This can be seen by writing \eqref{YAM} in terms of $\langle\phi(\mathbf{x})\rangle\propto\gamma(\mathbf{x})^{-(d-2)/2}$. The result is the saddle point equation for an $O(N)$ Landau-Ginzburg action at the upper critical dimension 
\cite{Galvani2021}. For $d<d_c$, it is possible to modify the Yamabe equation in order to account for the anomalous dimension of the field $\Delta_\phi= \frac{d-2+\eta}{2}$. Dimensional analysis then suggests that the exponents in~\eqref{YAM} be altered: the laplacian 
is then replaced by a fractional derivative, leading to the fractional Yamabe equation \cite{MarGonzalez2010}
\begin{equation}
(-\bigtriangleup)^{d/2-\Delta_{\phi}}\gamma_{(\Delta_{\phi})}(\mathbf{x})^{-\Delta_{\phi}}\propto \gamma_{(\Delta_{\phi})}(\mathbf{x})^{-d+\Delta_{\phi}}.
\label{FYE}
\end{equation}
The fractional laplacian $(-\bigtriangleup)^s$ in \eqref{FYE} is a nonlocal operator 
with many possible definitions \cite{Kwasnicki2017}, all equivalent for infinite system, but no longer compatible once boundaries are introduced. 
For our purposes, the fractional laplacian must be computed through an extension to a $d+1$-dimensional space that transforms
consistently under local scale changes \eqref{metrica}, as first introduced in \cite{Graham2003} for the compact case. The solution of this equation makes a different kind of  curvature constant: the fractional $Q$-curvature \cite{MarGonzalez2013}. See Appendix A for details. 

We see that the conformal factor $\gamma_{(\Delta_\phi)}$ now depends explicitly on 
$\Delta_{\phi}$, meaning that the metric depends on both the shape of the domain and on the dimension of the field. 
The idea is then to solve the fractional Yamabe equation in the considered bounded domain and use the conformal factor to find the order parameter profile. 
For the purposes of determining the scaling dimension $\Delta_{\phi}$, only the 
one-point function is needed, as seen in \cite{gori} for the $3d$ Ising model and 
in \cite{Galvani2021} for the $3d$ XY model.
Fitting the profile of the order parameter with numerical data will give us the value of the scaling dimension $\Delta_{\phi}$. For a spin model, the order parameter $\braket{\phi(\mathbf{x})}$ is the magnetization, while for percolation it can be extracted by a slicing procedure
performed on the giant cluster. 
As a case study, we choose 
the geometry of a slab $[0,L]\times\mathbb{R}^{d-1}$, where $\gamma(\mathbf{x})$ only depends on the transverse direction $x\in [0,L]$, and use the conformal factor to find the order parameter profile.

It is instructive to apply this method to $2d$ case, where exact results are known \cite{difrancesco}. Taking the $d\rightarrow 2$ limit to the Yamabe equation \eqref{YAM}, one gets the Liouville equation:
\begin{equation}
    (-\bigtriangleup)\log \gamma(\mathbf{x}) = -\gamma(\mathbf{x})^{-2}.
\end{equation}
In $2d$, a metric is entirely defined by its scalar curvature: this means that the solution of the Liouville equation also solves the fractional Yamabe equation for any $\Delta_{\phi}$~\cite{gori}. In particular, for a strip
of width $L$ for $x\in[0,L]$, 
$\gamma(x)=\frac L{\pi}\sin \frac{\pi x}{L}$. 

\section{Continuum percolation} The discussed approach could be applied to simulation of lattice percolation models at the critical point. However, 
since the critical geometry approach uses as input the continuous order parameter profiles at criticality, it is advantageous to extract these profiles from simulations performed directly with a continuum model.
So we use a model consisting of  
objects placed continuously in space \cite{Rintoul1997,Mertens2012,Gori2015}. The algorithm is straightforward: $d$-dimensional balls with 
unit radius are generated, one at a time. The center of each is picked randomly, with uniform probability within a 
slab. If the ball intersects another one, it is added to the cluster of the latter. If it intersects two or more balls belonging to different clusters, the clusters 
are merged. We stop adding objects once the product of their number and the relative volume of one object reaches the critical filling fraction $\eta_c$, which means we are at the critical point: measurements can then begin. 
Details of the simulation are found in the Appendix B. 

\begin{figure}
    \centering
    \includegraphics[width=.21\textwidth]{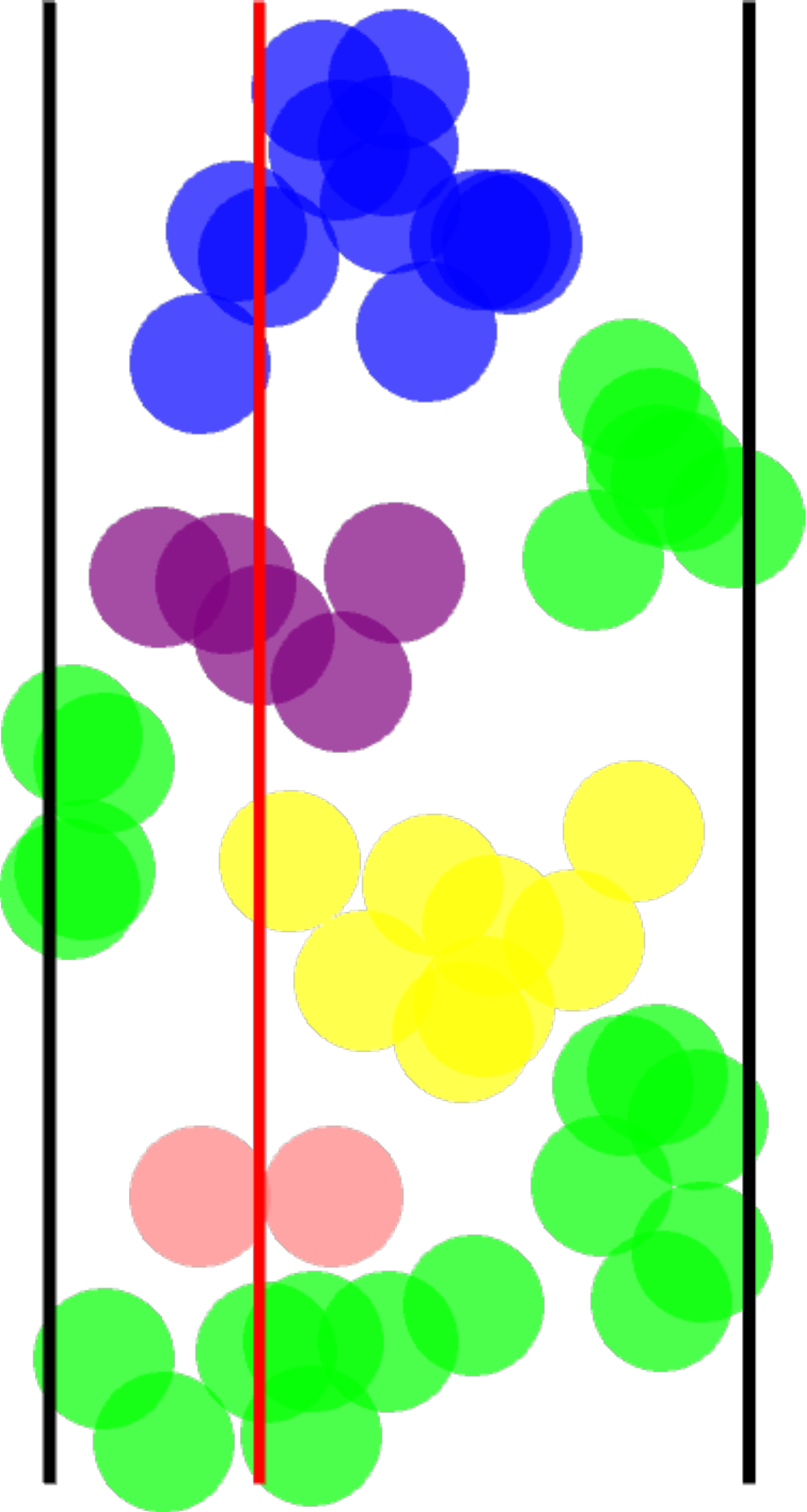}
    \hspace{.1cm}
    \includegraphics[width=.21\textwidth]{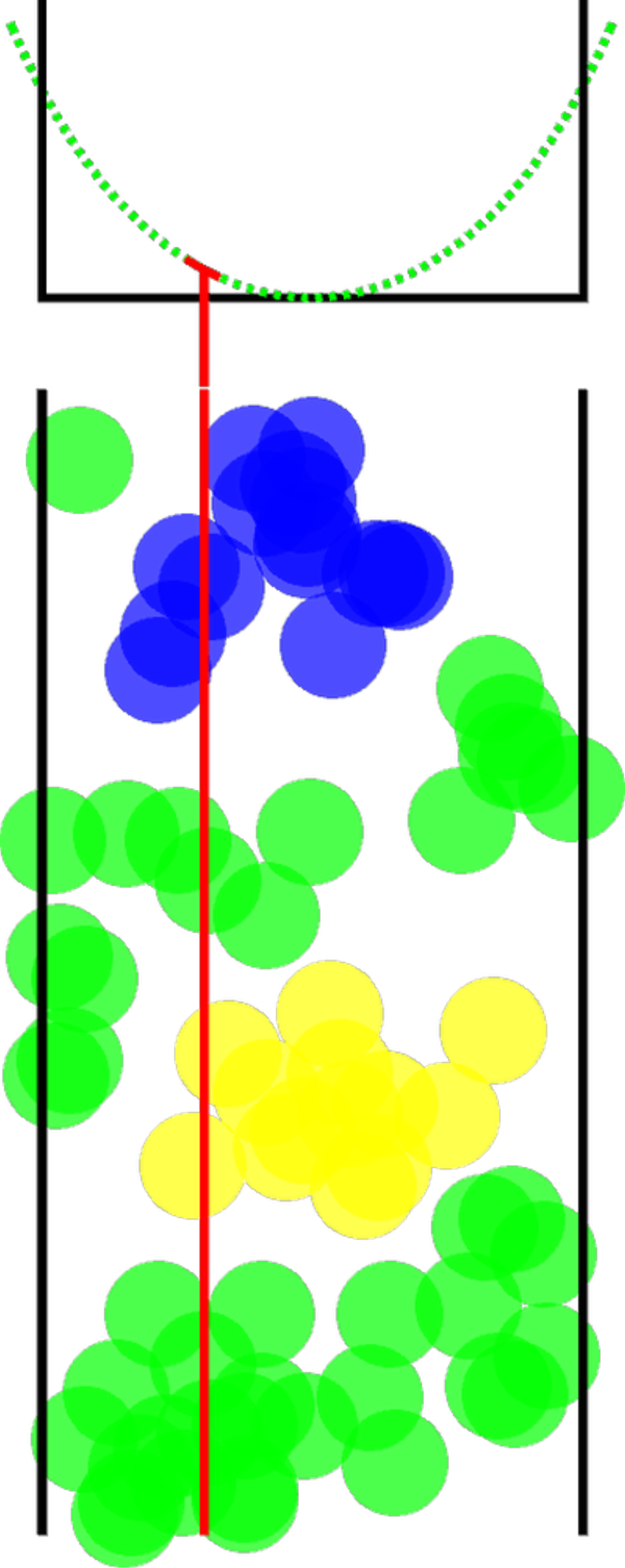}
    \caption{Continuum percolation in $2d$: circles on a strip. Fixed boundary conditions are enforced: when a circle intersects either boundary, it gets added to the large cluster (green). Left: $\eta<\eta_c$ --  
    Right: $\eta=\eta_c$. The segments obtained as intersection between the red line and the green balls contribute to the profile sketched above. The dashed green line illustrates the average on different realization, as shown by the data plotted in the next Fig.~\ref{profili2d}; it ends after the slab boundaries to illustrate the effect of the extrapolation length $a$.}
    \label{palle}
\end{figure}

Since divergences at the boundary are naturally produced, it is suitable for models with fixed boundary conditions. 
In the case of percolation, this means imposing from the start that the two boundaries belong to the same large (percolating) cluster. Any sphere which is added close to either boundary will then be added to the same cluster, as seen in Fig.~\ref{palle}. In lattice percolation, the one-point function or order parameter $\braket{\phi(x)}$, function of the distance from one boundary, is simply the probability that the lattice point $x$ belongs to the percolating cluster: 
this is measured by counting the number of lattice points, on a given plane parallel to the boundaries, which belong to the large cluster.

Here we see the main advantage of continuum percolation. The order parameter at a point $x$ is obtained by slicing the giant cluster with a plane, a distance $x$ from one boundary, and then measuring the total area (in $2d$, length) obtained as intersections between the objects and the plane. 
 This means that the profiles generated by the simulations are continuous themselves, lifting discretization effects  (finite size effects 
being of course still present).

\section{Check for $2d$ percolation}Before venturing into the $3d$ case, we checked that in $2d$ the method  gives a value for the exponent $\Delta_{\phi}$ in agreement with the analytical prediction.

The system is a strip of sizes (in units of the diameter of a sphere) $L$ in the transverse direction $x$ and $4L$ in the parallel direction $y$, along which periodic boundary conditions are imposed \footnote{Different ratios between the two sizes have been tested: increasing this ratio beyond 4 does not alter the following results.}.
We performed simulations for different values of $L$ 
 ranging from $L=16$ to $L=128$, in steps of $8$. The critical filling ratio we used is $\eta_c^{2d}= 1.12808737$ \cite{Mertens2012}~\footnote{
We checked that simulations performed at $\eta_c+\sigma_{\eta_c}$ and at $\eta_c-\sigma_{\eta_c}$ give profiles  indistinguishable within the error, where $\sigma_{\eta_c}$ is the error for $\eta_c$ given in \cite{Mertens2012}. In particular, for small sizes and especially in $2d$, where $\sigma_{\eta_c}$ is very small, varying the number of balls by just one changes the filling ratio from below $\eta_c-\sigma_{\eta_c}$ to above $\eta_c+\sigma_{\eta_c}$. This does not alter the profile and the subsequent $\Delta_{\phi}$, either in $2d$ or in $3d$.}
 
Continuous percolation allows us to get comparable results for different sizes, by measuring the order parameter $\braket{\phi(x)}$ across 
a fixed odd number $2 n-1$ of planes for every system size (even for small $L$), equally spaced throughout the slab. By symmetry, the values $x$ and $L-x$ have been averaged, meaning that each profile consists of $n$ points.

Using~\eqref{1p}, we fit the order parameter profiles with the function
\begin{equation}
\braket{\phi(x)}=\alpha \left[ L\, \gamma_{(\Delta_\phi)}\left( \frac x {1+a/L}\right)\right]^{-\Delta_{\phi}}\label{fitfunc},
\end{equation}
where the fit parameters are a multiplicative constant $\alpha$, the extrapolation length $a$ (accounting for the fact that the numerical profile does not diverge on the boundary~\cite{Cardy1996}) and the scaling dimension $\Delta_{\phi}$. 

\begin{figure}
    \centering
    \includegraphics[width=.6\columnwidth]{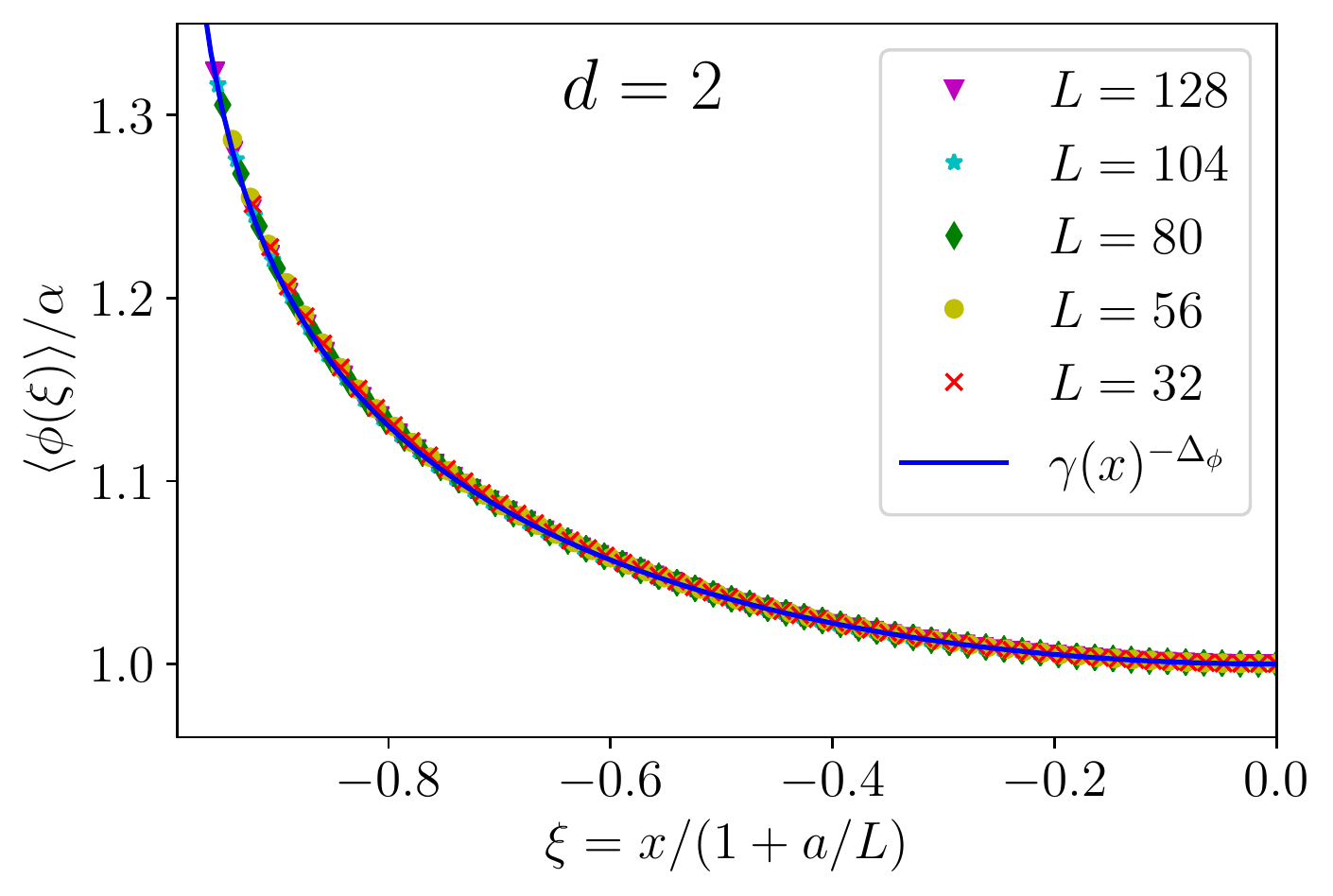}
    \caption{Order parameter profile for selected sizes. The $x$ coordinate has been rescaled for each size to take the extrapolation length into account.}
    \label{profili2d}
\end{figure}

Once the profiles for different sizes have been rescaled 
by multiplying each by $L^{-\Delta_{\phi}}$ and by plotting them as a function of $\xi=x/(1+a_L/L)$,
they are seen to collapse onto the same curve, as seen in Fig.~\ref{profili2d}, 
as expected at the critical point; 
each size gives us a value of $\Delta_{\phi}(L)$. To obtain the points plotted in Fig.~\ref{profili2d} 
an average on a few thousand (depending on $L$) realizations has been done. 

\begin{figure}
    \centering
    \includegraphics[width=.6\columnwidth]{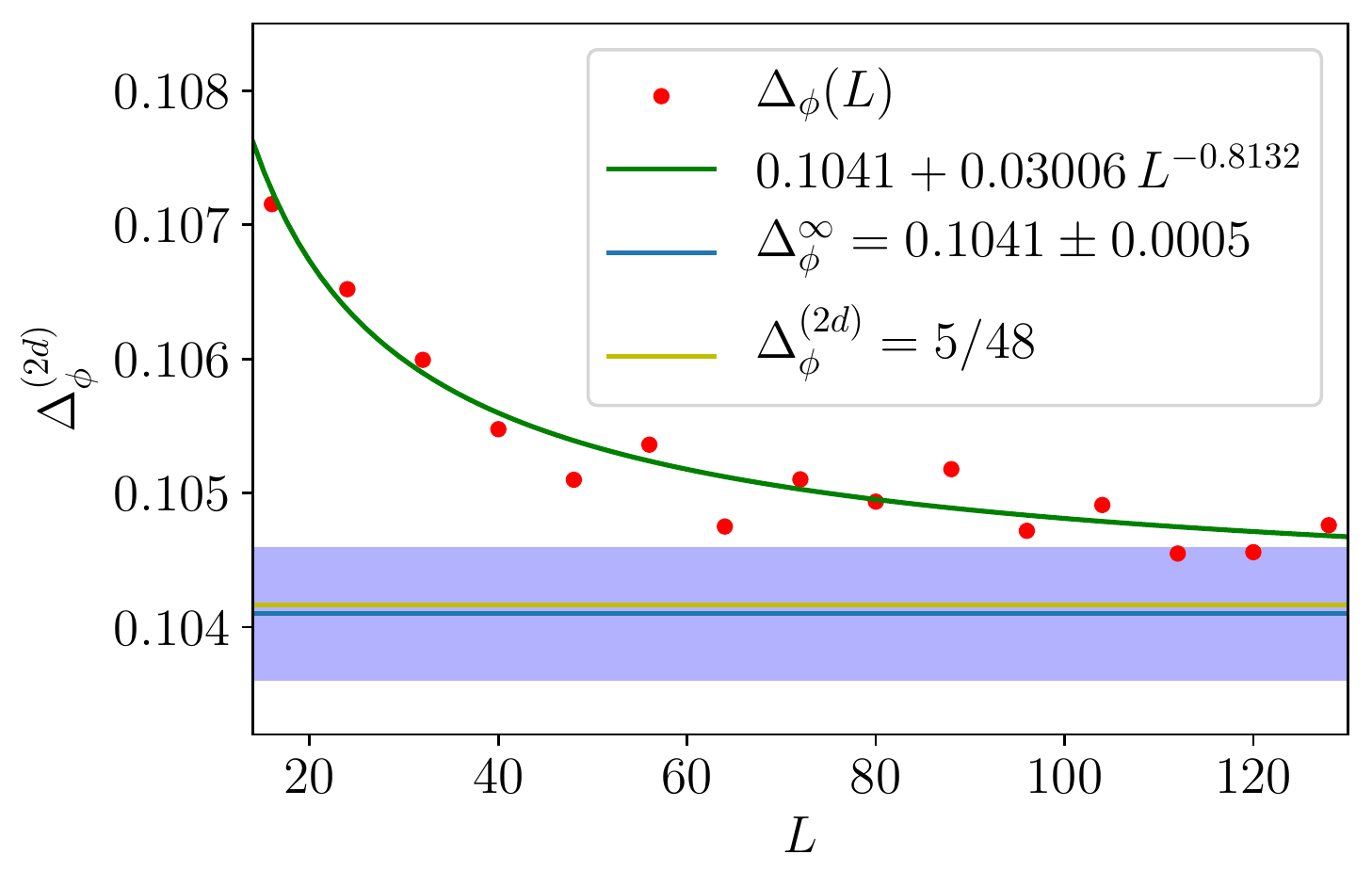}
    \caption{Fit results $\Delta_{\phi}(L)$ as the system size $L$ increases in $d=2$ (red dots). The  green line is the extrapolation fitting function~\eqref{powlawfit} while the continuous blue line is our best estimate $\Delta_{\phi}^\infty$ with the shading representing the error $\sigma$ on $\Delta_{\phi}^\infty$. The yellow line is the exact value. Despite using relatively small system sizes, 
    $\Delta_{\phi}^{\infty}$ 
    is convincingly close the exact value.}
    \label{varidelta2d}
\end{figure}

 What we can achieve in $3d$, we see in Fig.~\ref{varidelta2d} a decay of the fit parameter $\Delta_{\phi}^{(2d)}(L)$ as the size $L$ increases. To extrapolate the correct value, free of finite-size effects, we perform a fit in the form of a power law:
\begin{equation}\label{powlawfit}
    \Delta_{\phi}(L)=\frac {c}{L^k}+\Delta^{\infty}_{\phi},
\end{equation}
with $c,k,$ and $\Delta^{\infty}_{\phi}$ as fit parameters. This gives $\Delta_{\phi}^{\infty}=0.1041(5)$, which has to be compared with the exact value $\Delta_{\phi}^{(2d)}=5/48\approx 0.10417$. This is a good estimate, obtained with relatively small values of $L$ and 
with little numerical effort (once the numerical solution of the Yamabe equation is determined).

\section{Results for $3d$ percolation}The same can now be done for a $3d$ slab of sizes $L\times 4L \times 4L$; $L$ ranges from $16$ to $100$ in steps of $4$, and the critical filling fraction used is $\eta_c^{3d}=0.341888$, currently the most precise estimate \cite{Lorenz2001}.
An important difference with respect to the $2d$ case is the dependence of the fractional Yamabe equation on $\Delta_{\phi}$. We thus obtained a solution $\gamma_{(\Delta_{\phi})}$ that varies smoothly for $\Delta_{\phi}\in [0.46,0.5]$, which 
includes the correct value. We notice that $\gamma_{(\Delta_{\phi})}$ is almost constant in that range, which means that using the integer Yamabe equation (corresponding to $\Delta_{\phi}=1/2$) would be a reasonable initial approximation. See the Appendix A for further details.

\begin{figure}
    \centering
    \includegraphics[width=.6\columnwidth]{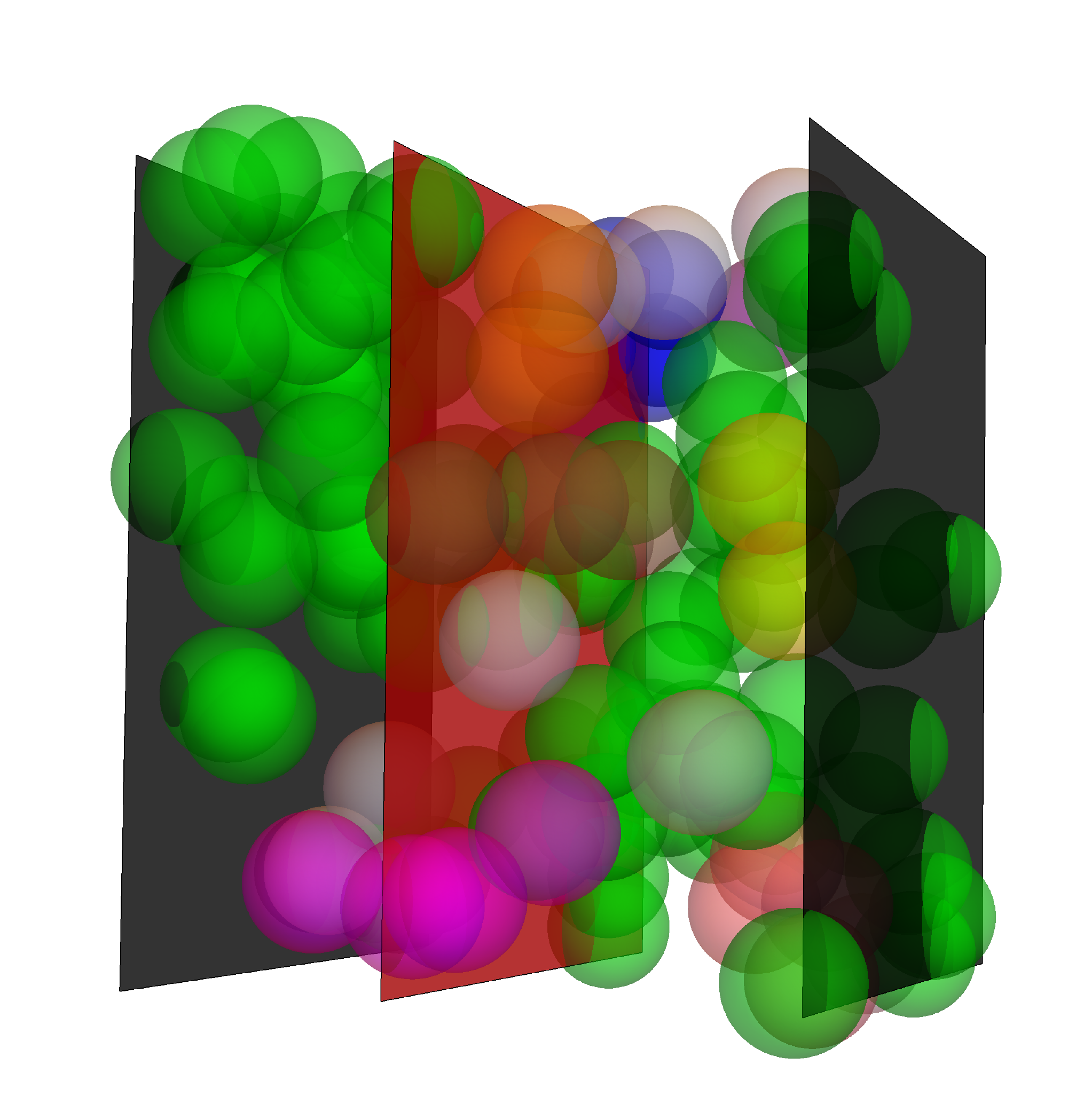}
    \caption{Slicing of the giant cluster (green spheres) with a plane (red). The black planes are the system boundaries: balls intersecting either belong to the giant cluster.}
    \label{fig:my_label}
\end{figure}

\begin{figure}
    \centering
    \includegraphics[width=.6\columnwidth]{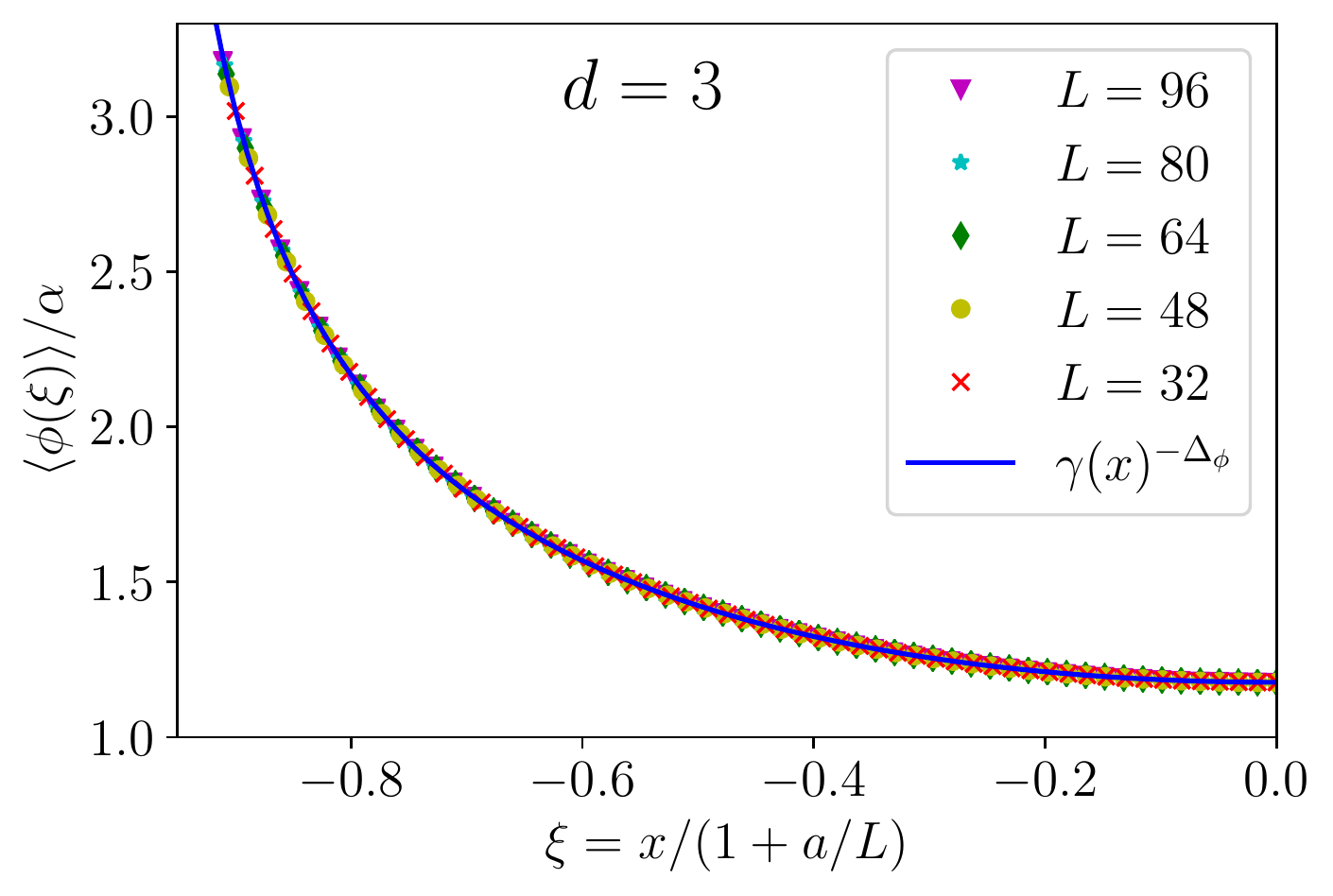}
    \caption{Order parameter profiles for $d=3$. We see a clear collapse onto the same curve, proving that we are at the critical point. The theoretical curve fits the data points accurately.}
    \label{profili}
\end{figure}

\begin{figure}\centering
    \includegraphics[width=.6\columnwidth]{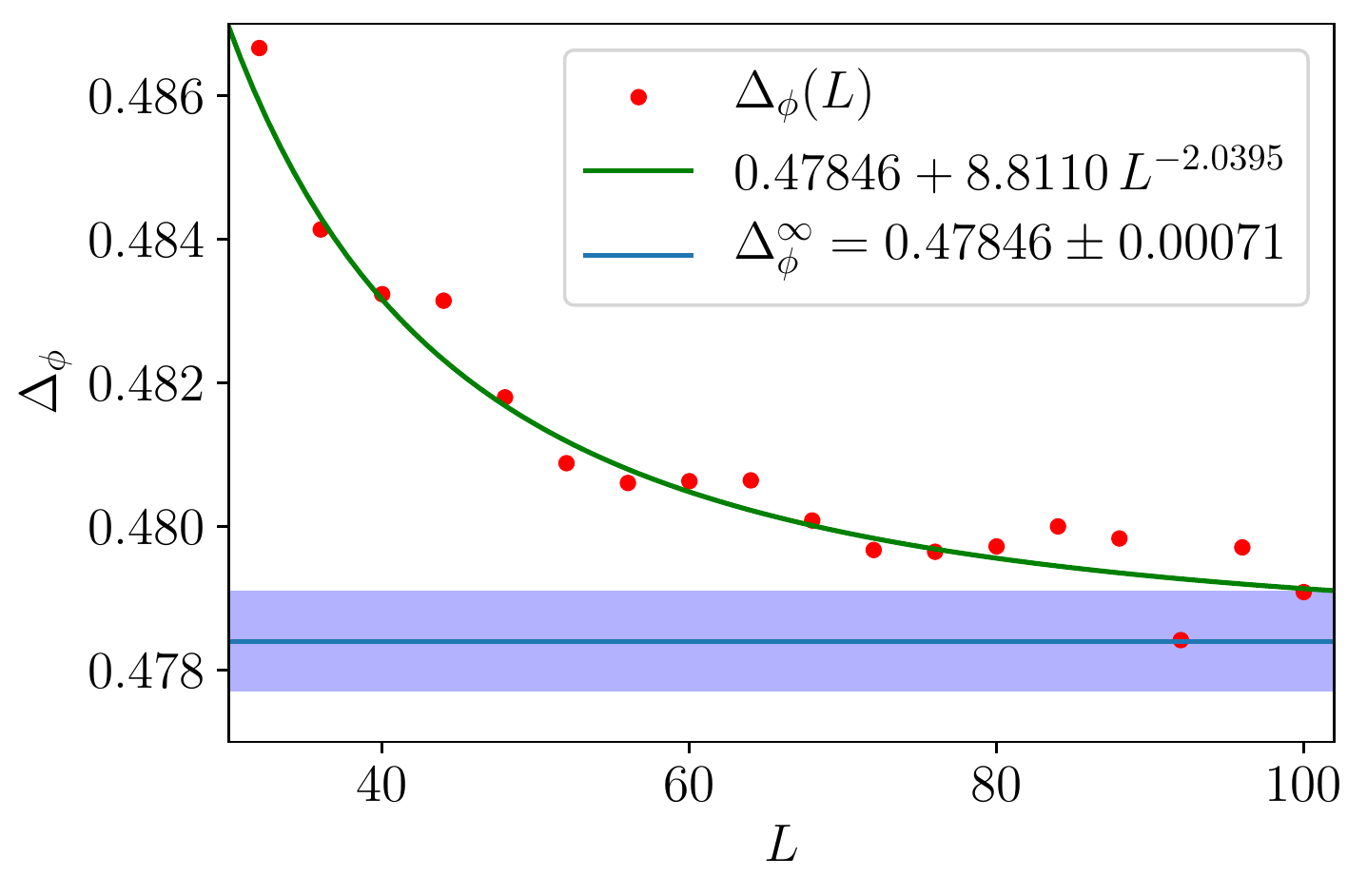}
    \caption{
Finite size estimates $\Delta_{\phi}(L)$ for the $d=3$ slab (red dots) as a function of $L$. The extrapolation curve~\eqref{powlawfit} is the continuous green line while the infinite size value $\Delta_{\phi}^{\infty}$ is the dashed blue line. The shaded area represents the error $\sigma$ on $\Delta_{\phi}^{\infty}$.}
    \label{varidelta}
\end{figure}

We obtain another clear collapse of the profiles in Fig.~\ref{profili}~\footnote{The same checks done in $2d$ for the role of $\sigma_{\eta_c}$, the ratio between the two sizes of the slab and the effect of the rounding of the number of balls needed to have the critical filling, have been repeated in $3d$.}. The profiles are then compared with the prediction for the profiles 
by our theory. The data are indeed described in an excellent way
by the fitting function~\eqref{fitfunc}.
For each system size values of $\Delta_{\phi}$ are obtained. Similarly to the $2d$ case a slight decay with $L$ (Fig.~\ref{varidelta}) is observed. 
By using a similar 
infinite-size extrapolation~\eqref{powlawfit} we obtain
our 
estimate for the scaling dimension:
\begin{equation}
    \Delta_{\phi}= 0.47846(71)  ,
\end{equation}
where the uncertainty is the statistical error on the fit parameter. From the definition
\begin{equation}
    \Delta_{\phi}=\frac{d-2+\eta}2,
\end{equation}
we get the corresponding anomalous dimension: \begin{equation}
\eta=-0.0431(14) .
\label{eta_f}
\end{equation}
This value is more precise than previous direct determinations of $\eta$ using other methods, listed in Table~\ref{tabelladelta}. 


\section{Scaling relations}Critical exponents are connected by well-known scaling relations \cite{Cardy1996}, which we can exploit to check the validity of our result. A scaling relation independent from the dimension $d$ and involving $\eta$ is 
\begin{equation}
    (2-\eta)\nu=\gamma;
\end{equation}
substituting our result, alongside $\gamma=1.805(20)$ \cite{Adler1990} and $\nu=0.8762(12)$ \cite{Xu2013}, we get $1.790(2)=1.805(20)$, meaning the equality is satisfied to one standard deviation.

Another class of relations between critical exponents is given by the hyperscaling relations \cite{Cardy1996}, where the dimensionality of the system enters explicitly. They are known to be violated above $d_c$. Tasaki derived a set of inequalities for the critical exponents of percolation \cite{Tasaki1987}. If these relations could be proven as equalities, then hyperscaling would in turn strictly hold. 
Out of the seven inequalities 
given in \cite{Tasaki1987}, two of them involve the exponent $\mu$, related to percolation with an ``external field". Four of them contain the rarely used exponents $\delta_r$ and $\Delta$ ($\Delta \equiv \beta+\gamma$). Another relation depends on $\nu$ but not on $\eta$. 
The remaining one, dependent on $d$, is
\begin{equation}
    (d-2+\eta)\nu -2\beta \geq 0.
    \label{hypereta}
\end{equation}


With our value of $\eta$ and $\beta=0.41(1)$ \cite{Sur1976}, one gets $(1-\eta)\nu-2\beta=0.018(22)$, so the left-hand-side of \eqref{hypereta} is compatible with $0$. 

If the hyperscaling equalities could be shown to hold, 
$\eta$ could be indirectly determined 
by measuring the fractal dimension $d_f$. Using then the relation $\eta=2+d-d_f$ and the value $d_f=2.52293(10)$ obtained in \cite{Xu2013}, 
one would obtain $\eta=-0.04586(20)$, which is compatible with our results within two standard deviations. 
Even with our result, the hyperscaling equality \eqref{hypereta} is satisfied to the second decimal digit, so it could be considered partly questionable to use it to compute $\eta$ to four decimal places. 
Another result for $\eta$ could be obtained through hyperscaling relations from the recent five-loops calculation \cite{fiveloop}, giving $\eta=-0.03(1)$.


\begin{table}
    \centering
    \begin{tabular}{c|c|c|c}
       Reference  & year & Method &$\eta$ \\
         \hline 
     Adler et al. \cite{Adler1990} & 1990  & Moment expansion & $-0.07(5)$ \\
     Lorenz \& Ziff  \cite{Lorenz1998} & 1998   & MC, bond percolation & $-0.046(8)$ \\ 
    Jan \& Stauffer
     \cite{Jan1998} & 1998 & MC, site percolation & $-0.059(9)$ \\
Gracey \cite{Gracey2015} & 2015 & 4-loop RG  & $-0.0470$ \\
    This work & 2021 & Critical geometry & $-0.0431(14)$
    \end{tabular}
    \caption{Comparison of the value of the anomalous dimension obtained  with various direct methods. Other results for $\eta$ making use of hyperscaling are reported in the main text.
    }
    \label{tabelladelta}
\end{table}

\section{Conclusions}We have constructed a purely geometric theory of percolation, the geometric model par excellence, at criticality. The spatial distribution of the giant cluster between the boundaries of the critical system has been linked to the solution of an equation for a metric having constant curvature in the considered domain, the fractional Yamabe equation. By using the solutions of this equation and results from the numerical simulation of continuum percolation, we  determined the anomalous dimension critical exponents $\eta$. Our results reproduce the known $2d$ result and compare favorably with previous determination of $\eta$ for $3d$ percolation, as seen in Table~\ref{tabelladelta}. 
The hyperscaling equalities with the obtained value of $\eta$ are shown to be satisfied within the error bars.

We have seen how the critical geometry approach can be used to determine the scaling dimension of a non-unitary model.
The validity of the critical geometric approach for one-point function here shown opens the possibility 
to study 
two-~and higher-point correlation functions, as done for the Ising and the XY model in \cite{gori,clock}. Further investigation is required to understand how this theory can describe fields other than the order parameter, correlations of different fields and boundary-condition-changing operators. The current theory can also be used to study percolation in $d=4,5,6$, as well as different geometries. 
\vspace{.5cm}\\
\textbf{Acknowledgements.} GG is supported by the Deutsche Forschungsgemeinschaft (DFG, German Research Foundation) under Germany’s Excellence Strategy EXC 2181/1 - 390900948 (the Heidelberg STRUCTURES Excellence Cluster). GG also acknowledges QSTAR for hospitality during completion of this work.

\appendix

\section{Solution of the fractional Yamabe equation}

\begin{figure}
    \centering
 \includegraphics[width=\columnwidth]{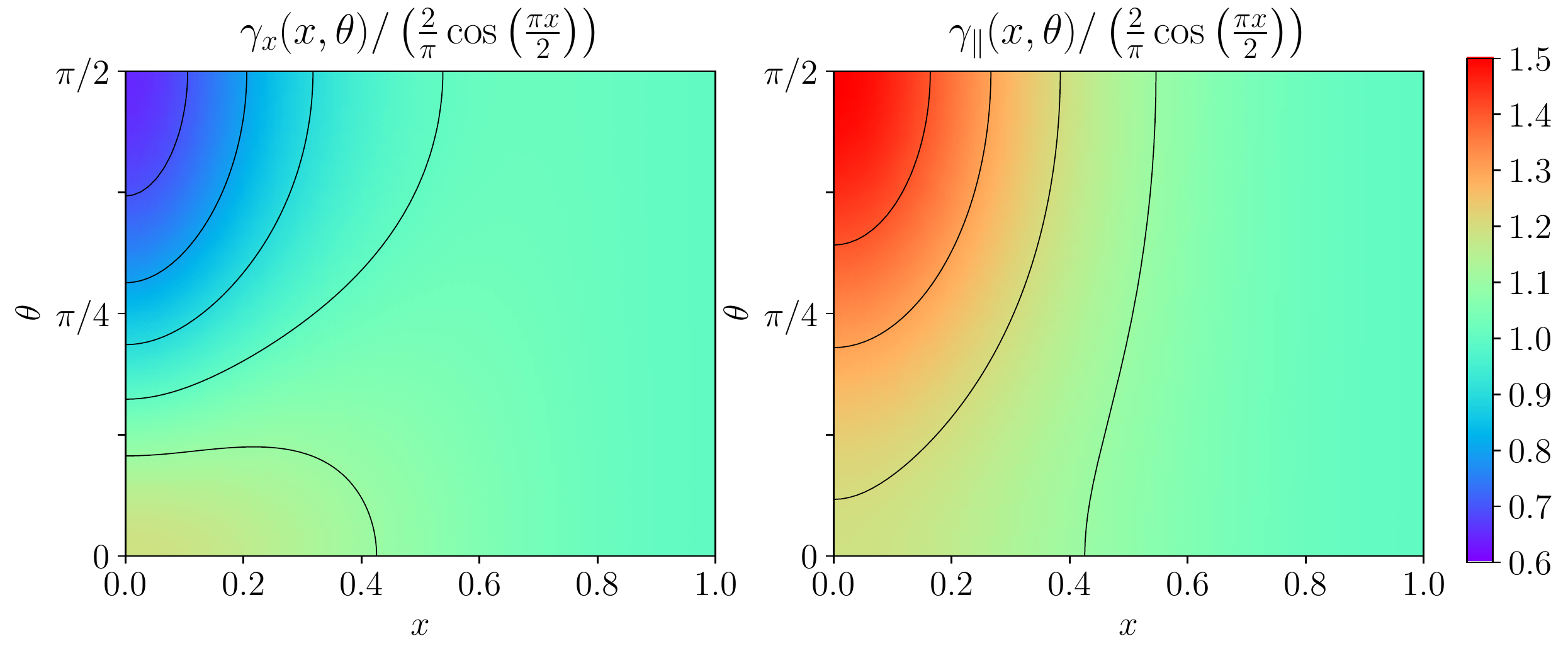}
 \caption{Functions defining the metric~\eqref{slab_g_plus}
 in the extension space above the slab. The 
 domain shown is $x\in[0,1]$ (the relevant transverse direction
 $x$ varies between $-1$ and $1$) and $\theta\in[0,\pi/2]$.
 The region $x\in[-1,0]$ can be obtained by reflection. To improve readability, the functions have been divided by $\frac{2}{\pi}\cos
 \left(\frac{\pi x}{2}\right)$. Figure taken from~\cite{gori}.
 }\label{fig_slab}
\end{figure}
The fractional Yamabe equation used
in the main text to obtain the order parameter profile
relies upon the consistent definition of a conformally 
covariant fractional Laplacian $(-\bigtriangleup)^s$ where
$s=\frac{d}{2}-\Delta_\phi$. Since this construction will be performed for
an arbitrary metric $g$ we introduce the
new symbol $\mathcal{L}^{(s)}_g$ for the fractional conformal laplacian 
while we use $(-\bigtriangleup)^s$ 
just when $g$ is the flat metric.
This is achieved by 
viewing the domain under consideration $\Omega$,
equipped with a metric $g$,
as the boundary of a $d+1$ dimensional 
manifold $X=[0,\pi/2] \times \Omega$  endowed with a 
metric $g_+$. In this enlarged space a solution
for an eigenvalue problem is searched; this technique,
inspired from AdS/CFT, has been mathematically 
introduced in~\cite{Graham2003} for a compact domain $\Omega$.
For a domain with boundary this procedure 
has to be adapted in the following way.
Denoting by $\theta\in [0,\pi/2]$ the extension
direction the metric should take the form
of a so-called cornered hyperbolic metric~\cite{McKeown2017}:
\begin{equation}\tag{S1}\label{CAH}
    g_+=\frac{1}{(\sin\theta)^2}(\mathrm{d}\theta^2 + g_\theta)
\end{equation}
where $g_{\theta=0}=g$ such that on the surface
$\theta=0$, to be identified with our original 
domain $\Omega$, we have $g_+\approx \theta^{-2}(\mathrm{d}\theta^2 + g)$ making it an asymptotically hyperbolic metric. Moreover $g_+$ has to satisfy the following conditions:
\begin{equation}\label{cornered}\tag{S2}
\begin{cases}
\mathrm{Ric}(g_+)+d\,g_+=0\\
\partial_\theta g_\theta|_{\theta=\pi/2}=0\\
g_\theta^{-1}|_{[0,\pi/2]\times\partial\Omega}\rightarrow 0
\end{cases}
\end{equation}
where $\mathrm{Ric}$ consistently refers to the Ricci
scalar curvature in $d+1$ dimensions.
For a discussion of the meaning of these conditions
the reader is referred to Appendix B of~\cite{gori}.
Once we have $g_+$ we set up the following 
eigenvalue (scattering) problem
for the Laplace-Beltrami operator $(-\bigtriangleup_{g_+}^{LB})$
relative to $g_+$:
\begin{equation}\label{PDE}\tag{S3}
 \begin{cases}
 (-\bigtriangleup_{g_+}^{LB}) U = \Delta_\phi (d-\Delta_\phi) U \\
 U = (\sin \theta)^{\Delta_\phi} F_I + (\sin \theta)^{d-\Delta_\phi} F_O
 \end{cases}
\end{equation}
for the function $U$ defined on $X$. The functions $F_I$ and 
$F_O$ are regular and give access to the
fractional laplacian as follows:
$\mathcal{L}^{(s)}_g f_I = c_s f_O$ 
where $c_s = 2^{2s}\frac{\Gamma (s)}{\Gamma(-s)}$, 
$f_I=F_I|_{\theta=0}$, and $f_O=F_O|_{\theta=0}$.

For the relevant three dimensional slab geometry
the cornered metric satisfying~\eqref{CAH} and~\eqref{cornered} has been obtained 
in the form:
\begin{equation}\label{slab_g_plus}\tag{S4}
 g_+ = (\sin \theta)^{-2} \left[\mathrm{d}\theta^2 + \mathrm{d}x^2/\gamma_x(x,\theta)^2 + (\mathrm{d}y^2+\mathrm{d}z^2)/\gamma_\parallel(x,\theta)^2\right].
\end{equation}
The functions $\gamma_x$ and $\gamma_\parallel$ are 
plotted in Fig.~\ref{fig_slab}.

Given $g_+$ the nonlinear eigenvalue
problem 
\begin{equation}\tag{S5}
(-\bigtriangleup)^{d/2-\Delta_{\phi}}\gamma_{(\Delta_{\phi})}(\mathbf{x})^{-\Delta_{\phi}}\propto \gamma_{(\Delta_{\phi})}(\mathbf{x})^{-d+\Delta_{\phi}}
\end{equation}
implying the solution of~\eqref{PDE} has been tackled numerically in an iterative
fashion yielding the desired solution 
for the fractional Yamabe problem in the 
slab. For further information the reader
is addressed to Appendix C of~\cite{gori}.

\begin{figure}
\centering
 \includegraphics[width=.45\textwidth]{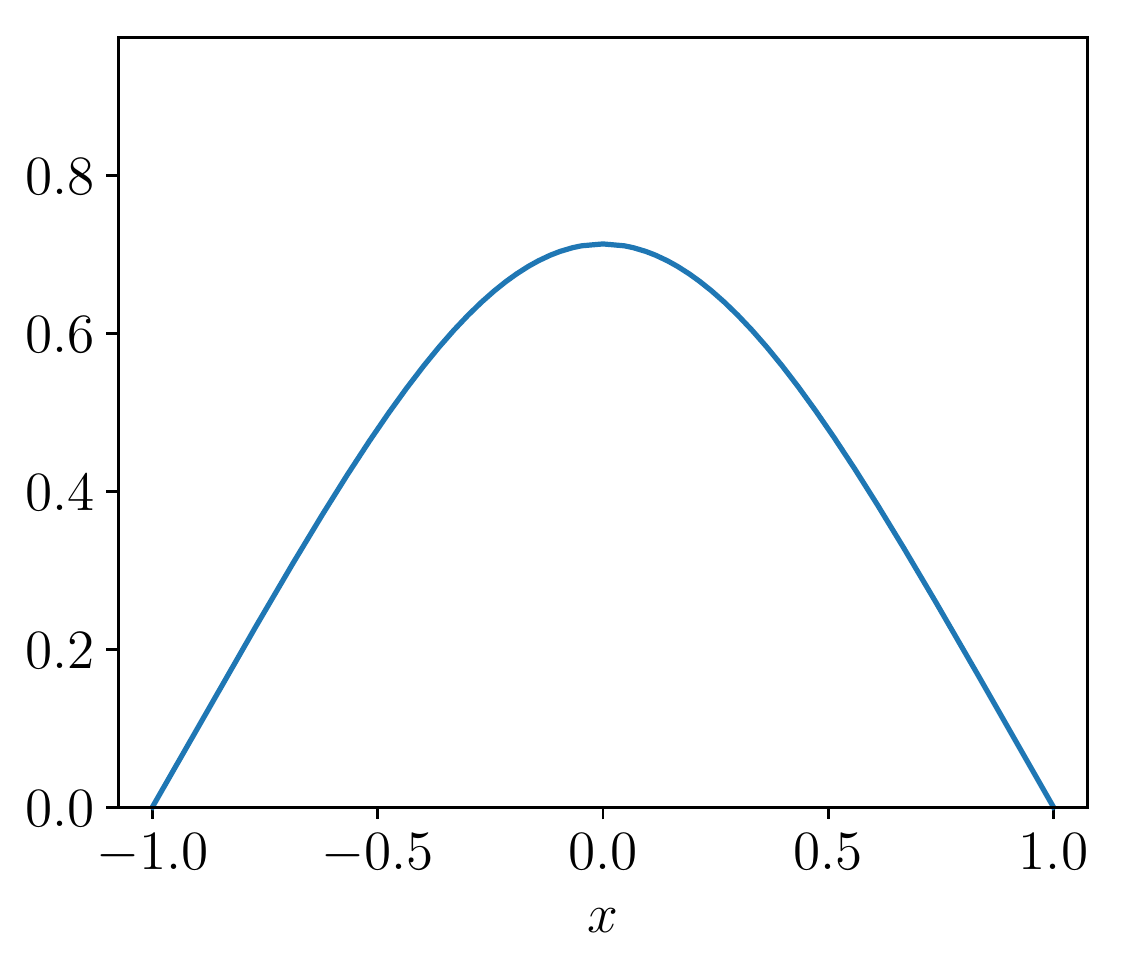}
 \includegraphics[width=.54\textwidth]{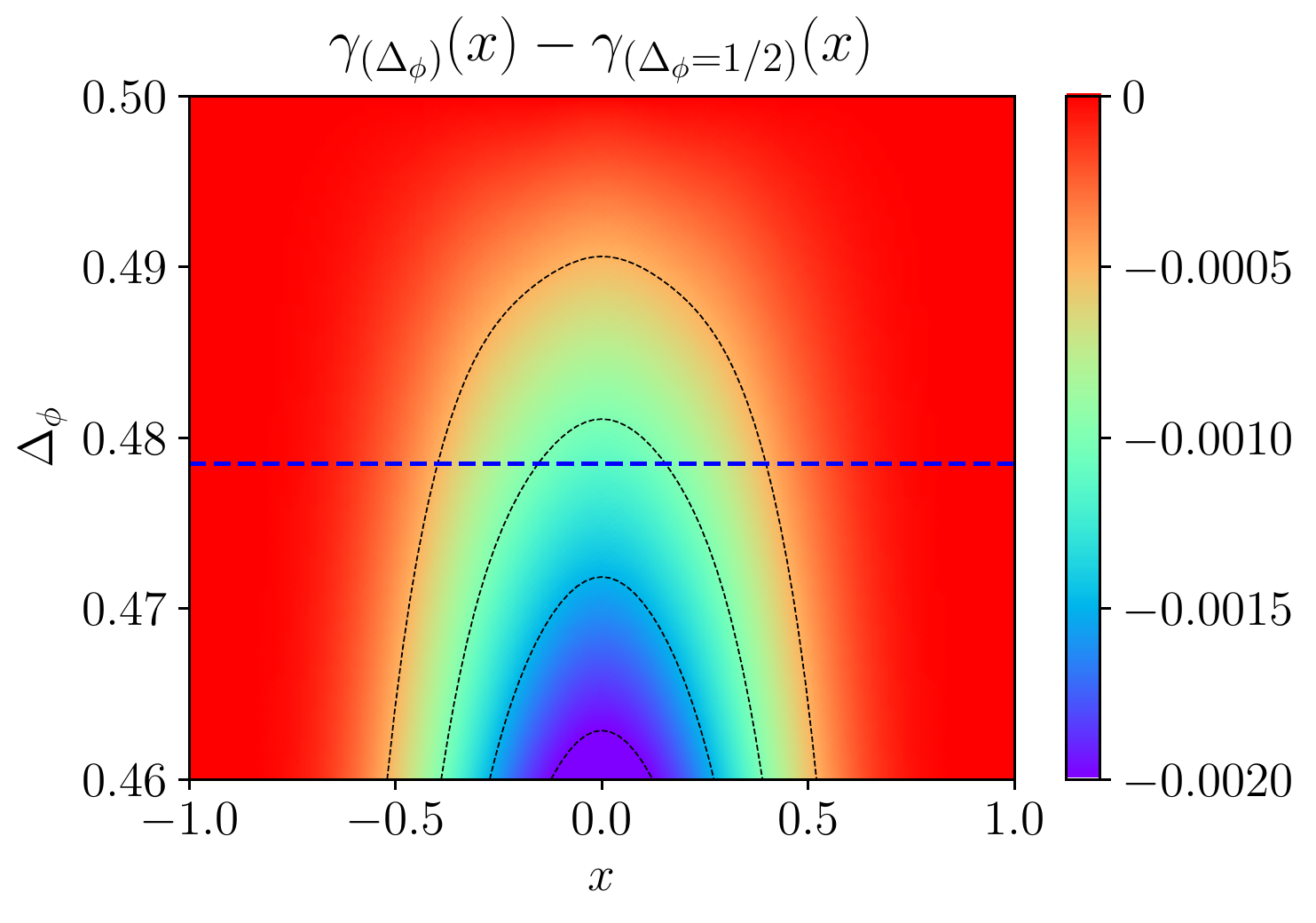}
 \caption{Solutions of the Yamabe problem in the three-dimensional slab domain ($-1<x<1$). Left: Integer Yamabe equation solution~\eqref{IYE3d}. Right: solutions of the Fractional Yamabe Equation 
 as $\Delta_\phi$ is varied in the range $[0.46,0.5]$. The plot 
 shows deviations from the $\Delta_\phi=1/2$ integer Yamabe solution (in the left panel). The blue dashed line is our best estimate as derived in the main text $\Delta_\phi=0.47846$.}\label{fig_IYEFYE}
\end{figure}

Results of this analysis are
shown in Figure~\ref{fig_IYEFYE}
for the range of anomalous dimensions
$\Delta_\phi \in [0.46,0.5]$
that is $\eta \in [-0.04,0]$ relevant for three-dimensional percolation.
In turn this means that we are solving a problem involving a laplacian raised to a power $s\approx 1.02$ \emph{greater} than 1. This is to be contrasted with the usual appearance of the fractional laplacian $(-\bigtriangleup)^s$ that has $s\in[0,1]$. Our numerical framework appears not to be affected by this fact.
Note that the small (in the $2\cdot 10^{-3}$ range) deviations from the integer Yamabe 
problem have been plotted.
The solution of the integer Yamabe problem
in three dimensions can be written explicitly (as derived in~\cite{Galvani2021}):
\begin{equation}\tag{S6}\label{IYE3d}
    \gamma_{(\Delta_\phi=1/2)}(x)=\frac{\sqrt{3}}{\omega} \wp\left(\omega (1+i\, \frac x{\sqrt{3}} ),\{0,1\}\right)
\end{equation}
where $\wp\left(z,\{0,1\}\right)$ is the equiharmonic case of the Weierstrass elliptic function with half period $\omega=\frac{\Gamma(1/3)^3}{4\pi}$.

\section{Details of the simulation}

For the sake of clarity, we will describe the algorithm used for the three-dimensional case; the two-dimensional version follows the same concept. 

The main difficulty of simulating continuum percolation, compared to the lattice variants \cite{Lorenz1998}, is to locate the objects that intersect the newly added one. To do this for the case of spheres, the entire slab has been divided in cubes of size equal to the diameter of a sphere. Since we take the diameter of the spheres to be 1, the number of these boxes will be $N=L\times 4 L \times 4 L$. Two matrices $C$ and $P$ are then introduced, with $N$ rows and variable number of columns, whose elements are themselves arrays: they will store, respectively, the  coordinates of the sphere centers and a pointer. A new sphere is added by generating the coordinates of its center, uniformly within the slab. From them, we determine to which box it belongs, say  the $n^{\text{th}}$ box, which already contained $k$ balls: an array containing the three coordinates is added to $C_{n,k}$, and we also set $P_{n,k}=(-1,0)$, to signify that the new sphere does not yet belong to any cluster. 
Then, we locate all the boxes that could contain spheres intersecting the newly added one: if box $n$ is not on a boundary, we have to check 27 boxes,  a $3\times3\times3$ grid centered in $n$. For each sphere in one of these boxes, we compute the distance between the two centers: if this is less than the sphere diameter, then an intersection has happened. Now we need to obtain the cluster to which the neighboring sphere belongs, and if it not the same as the cluster of the new sphere, the two will be merged. This is done by a ``union/find'' algorithm 
\cite{Newman2001}. The idea is to label each sphere so that it points to a sphere in the same cluster. The cluster can then be considered a tree, with various branches growing from one root. 
The first step is defining a find function: when it is fed the values representing a sphere, $(n,k)$, it looks at the values in $P_{n,k}=(n',k')$. If $n'$ is negative, by convention, it means that the $k'^{\text{th}}$ sphere in box $n'$ is the root of its cluster. Otherwise, the search continues, as we find the point that $(n',k')$ points to: $P_{n',k'}=(n'',k'')$. \\ In order to shorten the path for the next time the function is be called, once the root is found, the pointer of every sphere along the branch is changed so that it points directly to the root.
Next is the ``union'' part of the algorithm. Once the root of the intersecting sphere is known, if it is the same as the newly added sphere, nothing needs to be done. If the two roots differ, the smaller cluster must be included into the larger one, simply by changing the pointer of its root to the root of the larger cluster. Adding the smaller cluster to the larger one ensures that the average path to the root is shorter, but it might seem that additional effort is required to keep track of the cluster size. However, there is some convenient space to store this information that so far has not been used: the pointer of a root. While the pointer of every other sphere is the location of another sphere, so far we only specified that for a root site $(m,q)$, $P_{m,q}=(\alpha,\beta)$ with $\alpha<0$ to distinguish it from other sites. We can set $-\alpha$ to be equal to the number of sites in the cluster, and $\beta=0$ since it does not need to contain any information.
Now, when this cluster is merged with another one with root $(m',q')$ and $P_{m',q'}=(\alpha',0)$, assuming $|\alpha|>|\alpha'|$, we just have to set $P_{m',q'}=(m,q)$ and $P_{m,q}=(\alpha+\alpha',0)$. This links the smaller cluster to the larger one and updates the size of the latter. 

The entire process repeats by adding new balls until the critical filling ratio $\eta_c$ is reached. 
The main perk of this algorithm is that each union/find step takes an effectively constant computational time, i.e. it grows extremely slowly with system size. This means that the time to run the entire simulation is, for all practical purposes, simply proportional to the number of balls needed to reach the critical filling fraction. 

As previously mentioned, we want to implement fixed boundary conditions. To do so, we add a special object, which is adjacent to all the balls whose center is in a box on either boundary. This ensures that the balls in the first or last layer of boxes all belong to the same cluster, which is then the percolating cluster.


\section{Data analysis}

After obtaining the order parameter profiles, an additional step is needed before performing the fit. The points closest to the boundary are most affected by finite-size effects. Therefore, despite having smaller errors than the central points, a few of them have to be discarded. In order to determine how many to discard in an unbiased way, as well as to avoid a sharp distinction between discarded and included points, we introduce a window function $w(x)$. The weight of each point in the fit is given by the square of the ratio between this function and the error of that point. This function starts off from 0 at the boundary, ramps linearly to $1$ around a movable point $t$, and maintains the value $1$ up to the center of the slab.

To determine the location of the point $t$, we start from $t=-1$ (the boundary point) and gradually move towards $t=0$. For each value of $t$ we compute the $\chi^2$ of our data, and the corresponding p-value. We stop once the p-value reaches the reference value of $p=0.95$. 

\bibliography{bibGRAFFE.bib}

\end{document}